\documentclass{article}

\usepackage{amsmath}
\usepackage[pdftex]{graphicx}
\usepackage{float}
\usepackage{lscape}
\usepackage{verbatim}
\usepackage{setspace}
\usepackage{multirow}

\usepackage{todonotes}
\usepackage{changepage}
\usepackage{pdfpages}
\usepackage{rotating}
\usepackage{graphicx}
\usepackage{latexsym}
\usepackage{amssymb}
\usepackage{amsfonts}
\usepackage{longtable}
\usepackage{mathrsfs}
\usepackage{soul}
\usepackage{color}
\usepackage{epstopdf}
\usepackage{pbox}

\usepackage{url}

\definecolor{Gray}{gray}{0.85}

\def\href#1{\relax}

\title{Impact of magnetization and hyperfine field distribution on high magnetoelectric coupling strength in BaTiO$_3$–BiFeO$_3$ multilayers}
\author{Johanna K. Jochum$^{1}$, Michael Lorenz$^2$,  Haraldur P. Gunnlaugsson$^3$,\\ Christian Patzig$^4$, Thomas Höche$^4$, Marius Grundmann$^2$, \\ André Vantomme$^3$,Kristiaan Temst$^2$, Margriet J. Van Bael$^1$ \\ and Vera Lazenka$^3$}
\date{%
    $^1${KU Leuven, Laboratorium voor Vaste-Stoffysica en Magnetisme, Celestijnenlaan 200D, 3001 Leuven, Belgium}\\%
    $^2$ Felix-Bloch-Institut für Festk\"orperphysik, Universit\"at Leipzig, Linnéstraße 5, D-04103 Leipzig, Germany \\
    $^3$ KU Leuven, Instituut voor Kern en Stralingsfysica, Celestijnenlaan 200D, 3001 Leuven, Belgium \\
    $^4$ Fraunhofer-Institut für Mikrostruktur von Werkstoffen und Systemen (IMWS), Center for Applied Microstructure Diagnostics, Walter - H\"ulse - Straße 1, D-06120 Halle, Germany\\[5ex]%
    \today
}

\begin{document}

\maketitle                       

\begin{abstract}
Understanding the mechanisms of magnetoelectric (ME) coupling within multiferroic structures is paramount from a fundamental as well as an applied point of view. 
We report here that the magnetoelectric properties, as well as the magnetization, of BaTiO$_3$-BiFeO$_3$ superlattices can be tuned by varying the BiFeO$_3$ layer thickness. 
The magnetoelectric voltage coefficient ($\alpha_{ME}$) reaches its maximum of 60.2~V~cm$^{-1}$~Oe$^{-1}$ at 300~K, one of the highest values reported so far, for a sample with a BiFeO$_3$ thickness of 5~nm and a BaTiO$_3$ thickness of 10~nm. 
To gain deeper insight into the increased magnetoelectric coupling, and both the local and macroscopic magnetic properties, samples with varying BiFeO$_3$ thicknesses have been investigated. Correlations were established between the hyperfine field (HFF), the magnetoelectric voltage coefficient and the magnetization.
The possible mechanisms responsible for the strong magnetoelectric coupling are discussed.
\end{abstract}

\section{Introduction}

Magnetoelectric multiferroic heterostructures are of high interest due to their increased magnetoelectric (ME) coupling, as compared to the very few single phase magnetoelectrics such as BiFeO$_3$. 
In fact BiFeO$_3$ is the only known single phase multiferroic with both ferroelectric and antiferromagnetic ordering temperatures above room temperature (antiferromagnetic Néel temperature T$_N \approx$ 640~K and ferroelectric Curie temperature T$_C \approx$ 1100~K). 
Bulk BiFeO$_3$ is a G-type antiferromagnet with a cycloid spin structure with a period of 62~nm \cite{Sosnowska1982}. 
Several mechanisms exist in multiferroic heterostructures that can contribute to an increased ME coupling, such as interfacial strain coupling between the different components \cite{Thiele2007, Sando2013, Becher2015} as well as magnetic coupling created by combining a ferromagnet with an electrically tunable antiferromagnet to form an exchange biased system, such that the exchange bias can be switched using an electric voltage \cite{Fiebig2016, Scott2007, Bibes2008, Trassin2016}.
Out of the multiferroic composite structures investigated so far, thin film heterostructures such as multilayers and thin film composites show the strongest ME coupling \cite{Vaz2010, Wang2011, Jin2014}. 
Wang et al. \cite{Wang2011} found a direct magnetoelectric voltage coefficient of 52~Vcm$^{-1}$~Oe$^{-1}$ for metglas/piezofiber/metglas structures far off their electromechanical resonance frequency.
Dong et al \cite{Dong2006} found ME voltage coefficients ($\alpha_{ME}$) up to 22~Vcm$^{-1}$~Oe$^{-1}$ for magnetostrictive/piezofiber laminates.
We have recently reached values for $\alpha_{ME}$ as high as 49.7~Vcm$^{-1}$Oe$^{-1}$ in BaTiO$_3$-BiFeO$_3$ multilayers with atomically sharp interfaces \cite{Lorenz2016}. 
A clear dependency of $\alpha_{ME}$ on the oxygen pressure during pulsed laser deposition growth could be found in these multilayers.
In a previous study, we found that an increase in the number of BaTiO$_3$-BiFeO$_3$ double layers from 2 to 20, while keeping the double layer thickness constant, can lead to perpendicular magnetic anisotropy \cite{Lazenka2017} in BaTiO$_3$-BiFeO$_3$ multilayers, while $\alpha_{ME}$ only increased slightly from 16.4~Vcm$^{-1}$Oe$^{-1}$ to 28.8~Vcm$^{-1}$Oe$^{-1}$.
Here, we report a strong increase of the ME voltage coefficient when decreasing the BiFeO$_3$ layer thickness in BaTiO$_3$-BiFeO$_3$ multilayers from 50~nm to 5~nm while keeping the number of double layers and the BaTiO$_3$ layer thickness constant.
Using conversion electron Mössbauer spectroscopy (CEMS), interesting correlations of hyperfine (HF) field, magnetization, and magnetoelectric voltage coefficients could be established.


\section{Sample growth and structure}

\begin{figure}[H]
	\centering
		\includegraphics[width=0.90\textwidth]{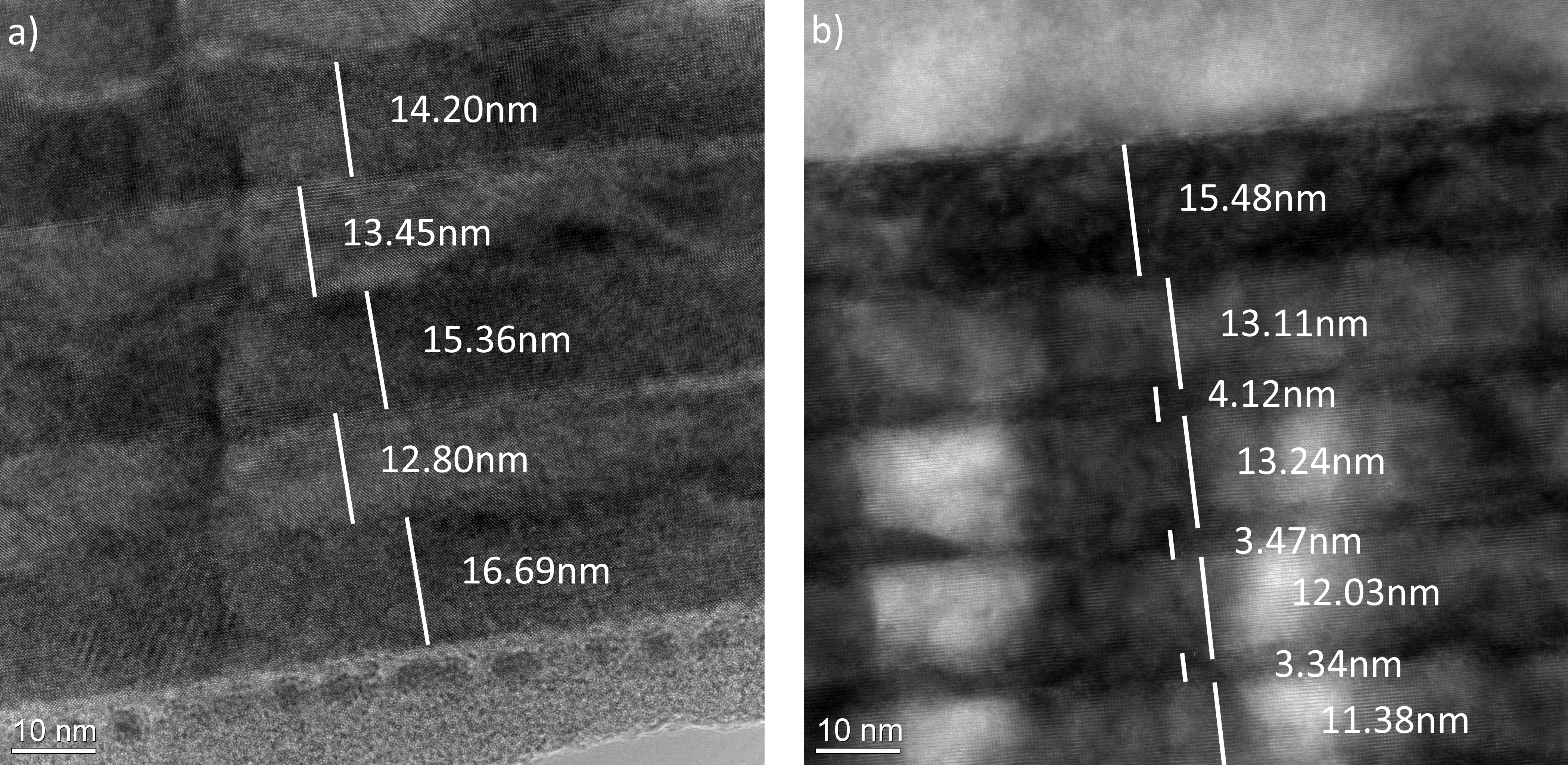}
	\caption{STEM profiles of the a) 15$\times$BTO-BFO(20) and b) 
15$\times$BTO-BFO(5) sample; the measured thicknesses of the BFO (darker) and BTO (lighter) layers have been indicated. The extracted BFO and BTO thicknesses lie in the range of 15-16~nm and 11-12~nm respectively for 15$\times$BTO-BFO(20) and 2-4~nm and 11-13~nm respectively for 15$\times$BTO-BFO(5).}
	\label{fig:STEM}
\end{figure}

Multilayers consisting of 15 BaTiO$_3$ - BiFeO$_3$ double layers with varying BiFeO$_3$ thickness (nominal thicknesses: 50~nm, 20~nm, 10~nm, 5~nm, hereafter referred to as 15$\times$BTO-BFO(50), 15$\times$BTO-BFO(20), 15$\times$BTO-BFO(10) and 15$\times$BTO-BFO(5), respectively) and a constant BaTiO$_3$ thickness of 10~nm have been grown by pulsed laser deposition from single phase BaTiO$_3$ and BiFeO$_3$ targets. The BiFeO$_3$ target contained 66.7\% isotopically enriched iron ($^{57}$Fe) in order to allow studying the local magnetic environment of the BiFeO$_3$ films using conversion electron Mösbauer spectroscopy (CEMS). The samples were grown simultaneously on SrTiO$_3$(001) as well as SrTiO$_3$:Nb(001) substrates, at an elevated temperature of 680$^{\circ}$C and an oxygen pressure of 0.25~mbar. More details on sample growth can be found in Reference \cite{Lorenz2014}. 
X-ray diffraction (XRD) $2\theta$--$\omega$ scans and reciprocal space maps (RSM) were measured at 300~K, utilizing a PANalytical X'Pert PRO MRD system with Cu K$_{\alpha}$ radiation using a parabolic mirror and a PIXcel$^{3D}$ array detector.
The double layer thicknesses (L) of the multilayers have been calculated from the $2\theta$--$\omega$ superlattice peaks using the relation $L = \frac{\lambda}{2\delta \theta \cos \theta_B}$, where $\lambda$ is the X-ray wavelength, $\delta \theta$ is the angular separation between two adjacent satellite peaks and $\theta_B$ is the Bragg angle of the zero-th order satellite peak. The calculated values can be found in Table \ref{tab:Structural}, for the 2$\theta$-$\omega$ scans the reader is referred to the \textcolor[rgb]{0,0,1}{\ref{Appendix}}. 
The observation that the multilayer peaks are not vertically aligned with the substrate peak in the RSM of the asymmetric ($\overline{1}$03) reflection implies that the multilayers are at least partially relaxed. From the symmetric (001) substrate peak the superlattice peaks can be clearly distinguished (see \textcolor[rgb]{0,0,1}{\ref{Appendix}}). The tilt mosaicity, which can be extracted from the horizontal broadening of the film fringes, is higher for the thicker samples than for the thinner samples.
Transmission electron microscopy (TEM), as well as high-angle annular dark-field scanning transmission electron microscopy (HAADF-STEM) and element distribution analysis by means of energy-dispersive X-ray spectroscopy (EDX)  have been performed on samples 15$\times$BTO-BFO(20) and 15$\times$BTO-BFO(5) using a state-of-the-art transmission electron microscope TITAN$^3$ 80-300 (FEI company). 
TEM micrographs showing the atomically sharp, layered structure of our multilayers can be seen in Figure \ref{fig:STEM}. The thicknesses of the BiFeO$_3$ and BaTiO$_3$ layers were extracted from these micrographs (see Table \ref{tab:Structural}). 
It can be seen that the BaTiO$_3$ layers are not perfectly flat, which leads to a modulation in thickness for the thin BiFeO$_3$ layers of sample 15$\times$BTO-BFO(5) from 2 to 4~nm. 
For the thicker BiFeO$_3$ layers in sample 15$\times$BTO-BFO(20), the modulation in BiFeO$_3$ thickness is lower (15-16~nm).

The local magnetic environment of the $^{57}$Fe atoms in the BiFeO$_3$ layers was probed using CEMS at room temperature and zero magnetic field. As in our previous work \cite{Lazenka2017}, the Mössbauer spectra were fitted using the ``hyperfine field (HFF) distribution model'' within the Vinda \cite{Gunnlaugsson2016} software. This model takes the possible coupling of isomer shift $\delta$ and hyperfine (HF) field $B_{HF}$ as well as quadrupole shift $\varepsilon$ and HF field into account: $\delta = \delta_0 + \frac{d\delta}{dB_{HF}}$ and $\varepsilon = \varepsilon_0 + \frac{d\varepsilon}{dB_{HF}}$. 

The macroscopic magnetic properties of our multilayers were investigated using SQUID (Superconducting QUantum Interference Device) magnetometry. Magnetization vs. field curves were recorded at 300~K in out of plane geometry with a LOT - Quantum Design MPMS 3 SQUID VSM. The magnetoelectric voltage coefficient was measured using a LOT - Quantum Design Physical Properties measurement System, making use of the relation: $\alpha_{ME} = \frac{dE}{dH}=\frac{1}{t}\frac{dV}{dH} \approx \frac{1}{t}\frac{V_{AC}}{H_{AC}}$, where t is the film thickness and $V_{AC}$ the AC voltage induced across the structures in response to a small applied AC magnetic field $H_{AC}$. $\alpha_{ME}$ was determined as a function of temperature and DC magnetic field at an AC magnetic field of 1~mT and a frequency of 1~kHz (for further details on the measurement method see reference \cite{Lazenka2015}).

\begin{table}[H]
	\centering
	\begin{adjustwidth}{-1.5cm}{}
\caption{BiFeO$_3$, BaTiO$_3$ and double layer thicknesses as determined from XRD and TEM; the nominal values (which are used throughout the manuscript for the sample description) are shown as well.}
		\begin{tabular}{|c | c | c |c | c | c | c |}
		\cline{1-7}
		 \multicolumn{1}{ |c  }{\multirow{2}{*}{Samples} } & \multicolumn{3}{ |c| }{nominal thicknesses}  & thickness, XRD & \multicolumn{2}{ |c| }{thickness, STEM} \\ 
		\cline{2-7}
			& \pbox{2.5cm}{BiFeO$_3$}	& \pbox{2cm}{BaTiO$_3$} & \pbox{2cm}{double layer} &	\pbox{2cm}{double layer}	& \pbox{2.5cm}{BiFeO$_3$}	& \pbox{2.5cm}{BaTiO$_3$}\\ 
			& [nm] & [nm] & [nm] & [nm] & [nm] & [nm] \\
		\cline{1-7}		
		15$\times$BTO-BFO(50) & 50	&	10 &	60	& 49 $\pm$ 2	&  $-$	& $-$ \\
		\cline{1-7}
		15$\times$BTO-BFO(20) & 20	&	10 &	30  &	26.3 $\pm$ 0.8	& 15-16 &	11-12 \\
		\cline{1-7}
		15$\times$BTO-BFO(10) & 10 &	 10	& 20 & 19.5 $\pm$ 0.2 & $-$ &	$-$ \\
		\cline{1-7}
		15$\times$BTO-BFO(5) & 5 &	10	& 15 &	15.3 $\pm$ 0.7	& 2-4 & 11-13 \\
		\cline{1-7}
		\end{tabular}
	\label{tab:Structural}
	\end{adjustwidth}
\end{table}

\section{Multiferroic properties}

\begin{figure}[H]
	\centering
		\includegraphics[width=0.90\textwidth]{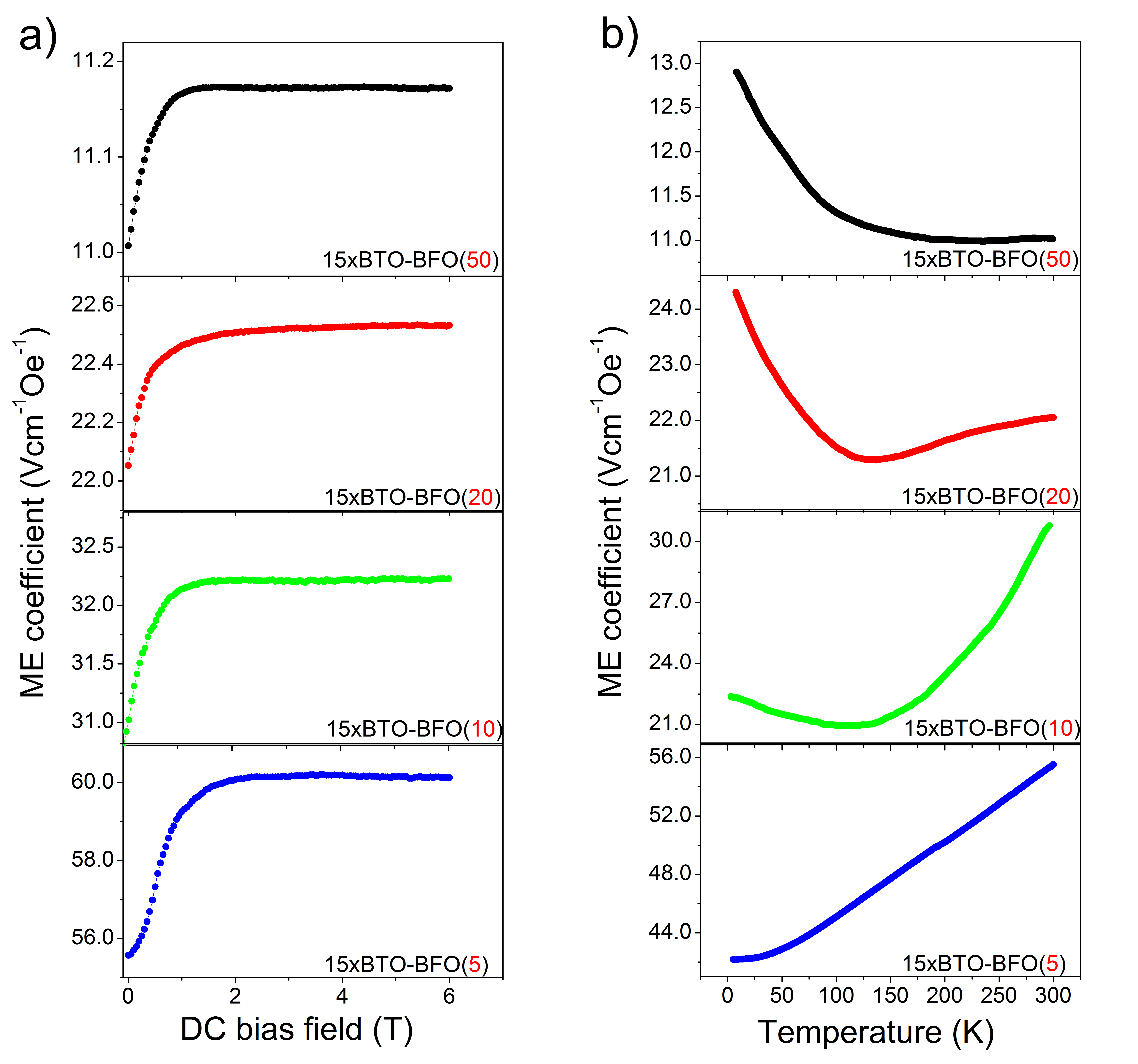}
	\caption{Magnetoelectric voltage coefficient of the multilayers as a) a function of magnetic bias field (at 300~K) and b) a function of temperature, for the indicated superlattice structures, see Table \ref{tab:Structural}.}
	\label{fig:ME_coefficient}
\end{figure}

The magnetoelectric voltage coefficient has been measured as a function of temperature and of DC bias field (see Figure \ref{fig:ME_coefficient}). 
$\alpha_{ME}$ reaches its maximum, off - resonance, value of 60.2~V~cm$^{-1}$~Oe$^{-1}$ for sample 15$\times$BTO-BFO(5) at a bias field of 2~T. 
This value of $\alpha_{ME}$ is, to the best of our knowledge, among the highest recorded in literature so far. 
The values for the other samples lie noticeably lower with 32~V~cm$^{-1}$~Oe$^{-1}$, 22.5~V~cm$^{-1}$~Oe$^{-1}$ and 11.2~V~cm$^{-1}$~Oe$^{-1}$ for samples 15$\times$BTO-BFO(10), 15$\times$BTO-BFO(20) and 15$\times$BTO-BFO(50), respectively.
The high values for $\alpha_{ME}$ in BiFeO$_3$ and BiFeO$_3$ heterostructures (discussed in our previous work \cite{Lorenz2014, Lorenz2015}) have been theoretically confirmed by Popkov et al. \cite{Popkov2016}.
They describe the rotation of the oxygen octahedra \cite{Lorenz2015} with the antiferrodistortion vector $\Omega$ and show that the linear ME effect depends on the derivative of $\Omega$ with respect to the external field.
This leads to giant values of $\alpha_{ME}$ at so called critical fields, where a reorientation of the antiferrodistortion vector occurs, and its derivatives approaches critical behaviour.\\
\vspace{\baselineskip}
While the tendency of the field dependence of the magnetoelectric voltage coefficient at 300~K is the same for the four samples, there is a significant difference in temperature dependence. 

The dependence of $\alpha_{ME}$ on temperature changes continuously with decreasing BiFeO$_3$ layer thickness. Sample 15$\times$BTO-BFO(50) shows a decrease of $\alpha_{ME}$ with increasing temperature as has been observed for bulk BiFeO$_3$ \cite{Modarresi2016} and for BaTiO$_3$-BiFeO$_3$ composite films \cite{Lorenz2016a, Lorenz2014}. 
For the samples with BiFeO$_3$ thicknesses of 20~nm and 10~nm, two different regimes can be distinguished in the temperature dependency of $\alpha_{ME}$: 
T<T$_{min}$ (T$_{min}$ corresponds to the temperature where $\alpha_{ME}$ shows the lowest value) where $\alpha_{ME}$ decreases with increasing temperature, and T>T$_{min}$ where $\alpha_{ME}$ increases with increasing temperature. 
This temperature dependence of $\alpha_{ME}$ was also observed in our previous work \cite{Lorenz2016, Lorenz2015}.
With further decrease in BiFeO$_3$ thickness, the temperature dependence of $\alpha_{ME}$ changes further, showing an increase of $\alpha_{ME}$ with increasing temperature for sample 15$\times$BTO-BFO(5). 
This change in temperature dependence as a function of BiFeO$_3$ layer thickness points towards a change in ME coupling mechanism as a function of the BiFeO$_3$ layer thickness.
Both the magnetostrictive and piezoelectric coefficients increase with increasing temperature \cite{Lorenz2016a}. It is therefore evident to conclude that a magnetoelectric multilayer dominated by the coupling of piezoelectric and magnetostrictive interaction via strain would show the same temperature dependence.
Therefore, it seems that there is a change from strain mediated coupling in sample 15$\times$BTO-BFO(5) to a regime where the multilayers show strain mediated coupling above a certain temperature T$_{min}$ while obeying a different coupling mechanism below T$_{min}$.
A possibility for such a mechanism is charge transfer across the interfaces \cite{Lorenz2016a, Spurgeon2014}. 
It has been suggested before that this change in the magnetoelectric coupling mechanism might be induced by a phase change in the BaTiO$_3$ layer\cite{Lorenz2015}. 
Bulk BaTiO$_3$ undergoes two structural phase transitions in the measured temperature range. The first transition, from tetragonal to orthorhombic, occurs at 273~K. The orthorhombic phase changes into a rhombohedral phase at 183~K \cite{Kwei1993}. The ferroelectric properties of the different phases of BaTiO$_3$ are strongly connected to the displacement of the Ti ions along the different faces of the oxygen octahedra, which leads to different polarization directions for the different structures. 
It is well-known that structural phase transitions can be shifted to either higher or lower temperatures in thin films by pseudomorphic growth on a substrate with a sufficiently small lattice mismatch \cite{Brune1997}. The lattice parameters of SrTiO$_3$, BaTiO$_3$ and BiFeO$_3$are close enough to each other to allow pseudomorphic growth. (STO, cubic: a = 0.3905~nm (JCPDS 84-0444); BaTiO$_3$, tetragonal: a = 0.39945~nm, c = 0.40335~nm (JCPDS 83-1880); BiFeO$_3$, rhombohedral: a = 0.3962~nm (JCPDS 73-0548)) 
As can be seen from the ($\overline{1}$03) RSM (see Figures S5 - S8 in the appendix), the layers are not lattice-matched with the substrate, however the in-plane lattice parameters of the BiFeO$_3$ and BaTiO$_3$ layers are matched to each other \cite{Lorenz2016}, leading to small in-plane strains (up to about 1\%). A possible explanation for the discussed change in the temperature dependence of $\alpha_{ME}$ might therefore be the onset of the orthorhombic-to-rhombohedral phase transition in the BaTiO$_3$ layers, which occurs at different temperatures for different BiFeO$_3$-BaTiO$_3$ layer thicknesses and repetitions.

\begin{figure}[H]
	\centering
		\includegraphics[width=0.90\textwidth]{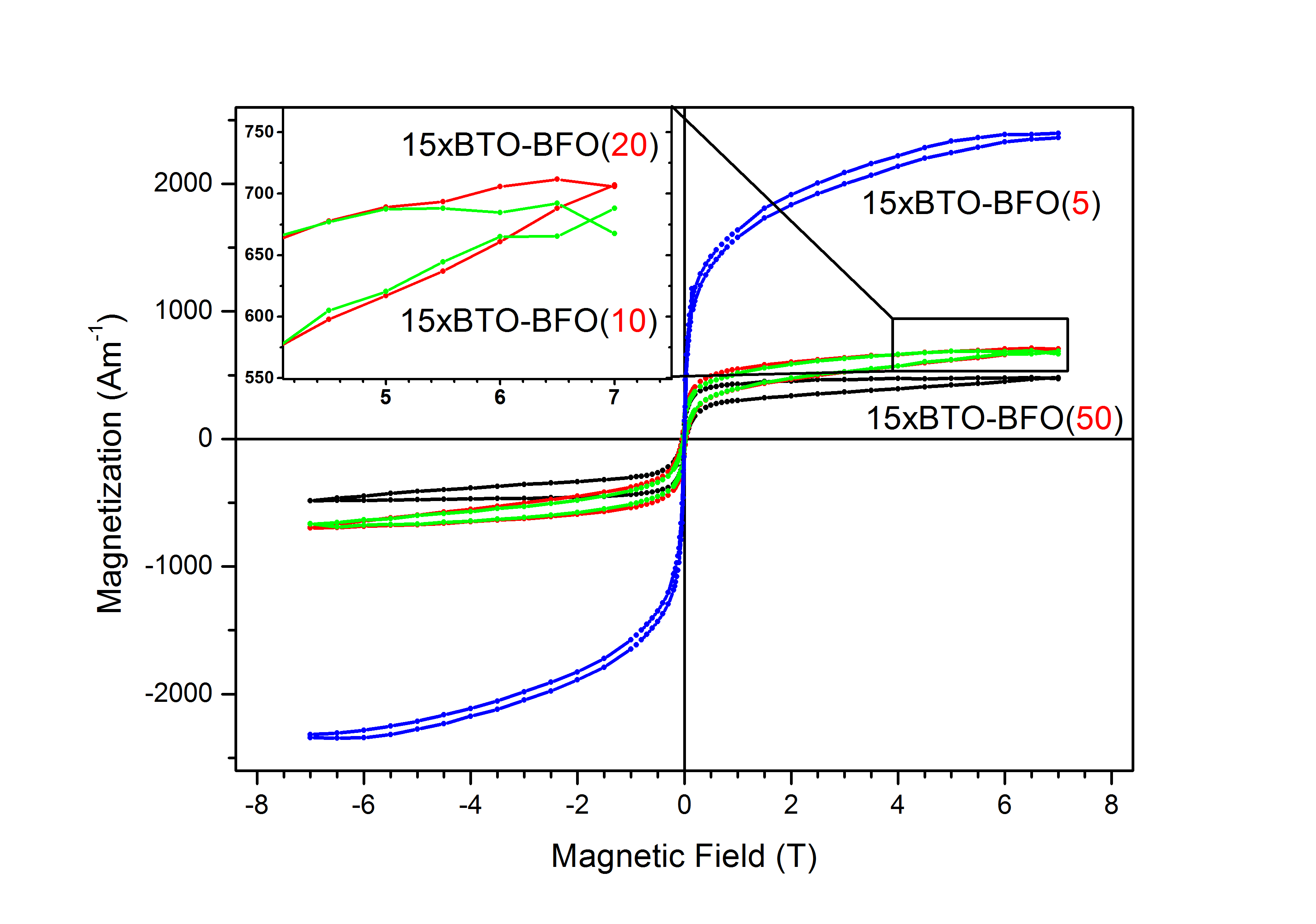}
	\caption{Magnetic hysteresis loop measurement of the multilayers in out-of-plane geometry taken at 300~K. The data have been normalized to the multilayer volume. The inset shows a zoom into the highlighted rectangle to show the curves of 15xBTO-BFO(20) and 15xBTO-BFO(10) more clearly.}
	\label{fig:OOP_straw_comparison}
\end{figure}

To find an explanation for the significantly increased ME voltage coefficient for 15$\times$BTO-BFO(5) we investigated the macroscopic and local magnetic properties of our multilayers. 
SQUID magnetometry measurements were done in out-of-plane configuration, the same geometry as was used for the ME voltage coefficient measurements. 
As can be seen in Figure \ref{fig:OOP_straw_comparison} the (volume) magnetization (in Am$^{-1}$) of 15$\times$BTO-BFO(5) exceeds the (volume) magnetization of the other three samples by more than a factor of two, following the same trend as the ME voltage coefficient when comparing the results as a function of BiFeO$_3$ thickness. The same tendencies for the dependence of magnetization as well as ME voltage coefficient on BiFeO$_3$ thickness have been observed by Kotnala et al. \cite{Kotnala2015} for BaTiO$_3$/BiFeO$_3$/BaTiO$_3$ trilayers. 
Figure 3 further shows that none of the samples is saturated at a field of 7~T. For samples 15$\times$BTO-BFO(50), 15$\times$BTO-BFO(20) and 15$\times$BTO-BFO(10) coercive fields of approximately 35~mT were found which is comparable to what has been found by other groups \cite{Yang2009, Kuo2016}. The coercivity almost disappears for 15$\times$BTO-BFO(5).

\begin{figure}[H]
	\centering
		\includegraphics[width=0.90\textwidth]{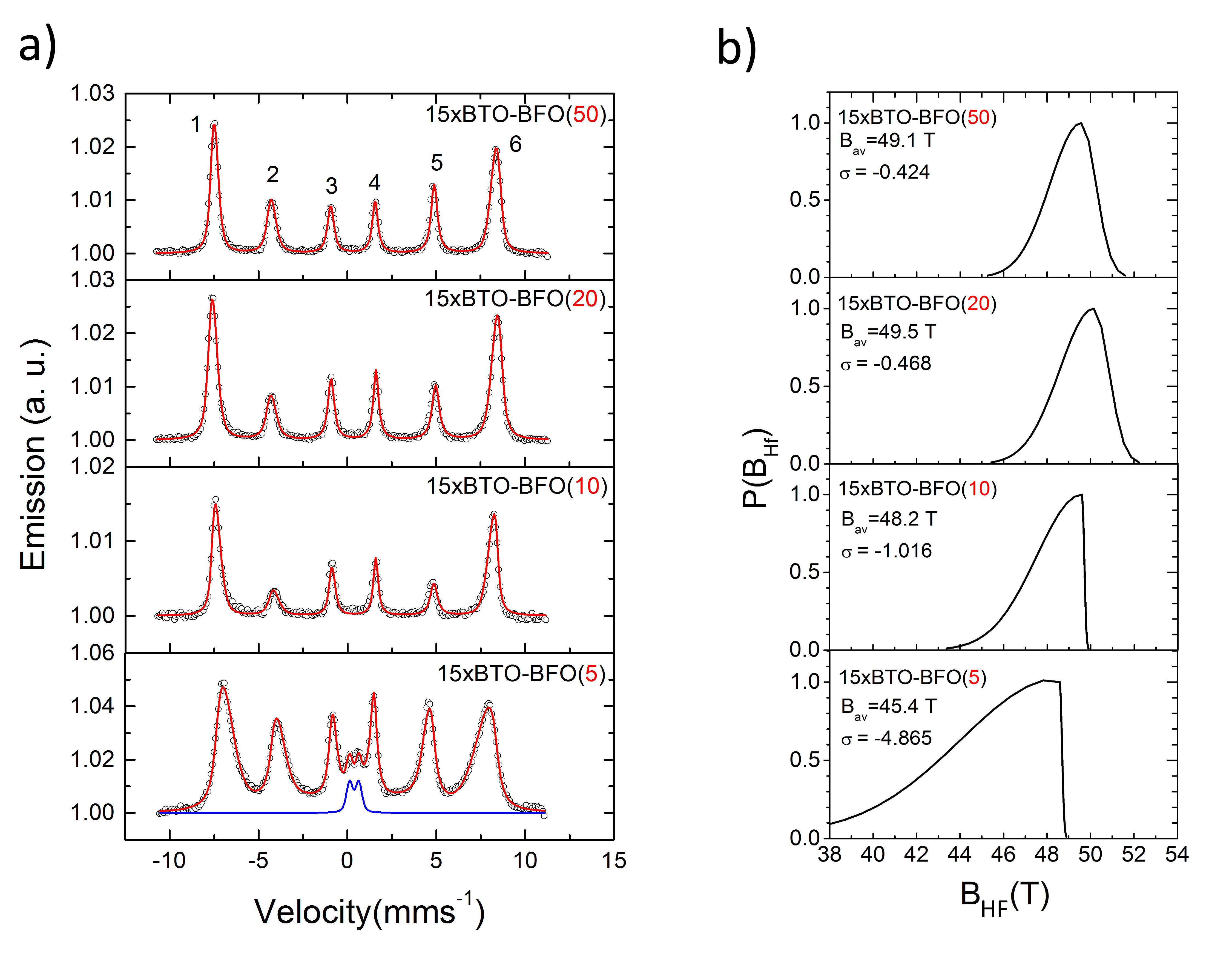}
	\caption{a) Conversion electron Mössbauer spectroscopy spectra of the multilayers recorded at 300~K. The white circles correspond to the data, whereas the red lines represent the fits. The blue line for sample 15$\times$BTO-BFO(5) corresponds to the non magnetic doublet needed to accurately fit the data; b) Hyperfine field distributions of the multilayers; the average hyperfine fields of the different samples are shown; furthermore, the skewness value $\sigma$, a measure of the distribution's asymmetry is indicated.}
	\label{fig:Moessbauer}
\end{figure}

\begin{figure}[H]
	\centering
		\includegraphics[width=0.90\textwidth]{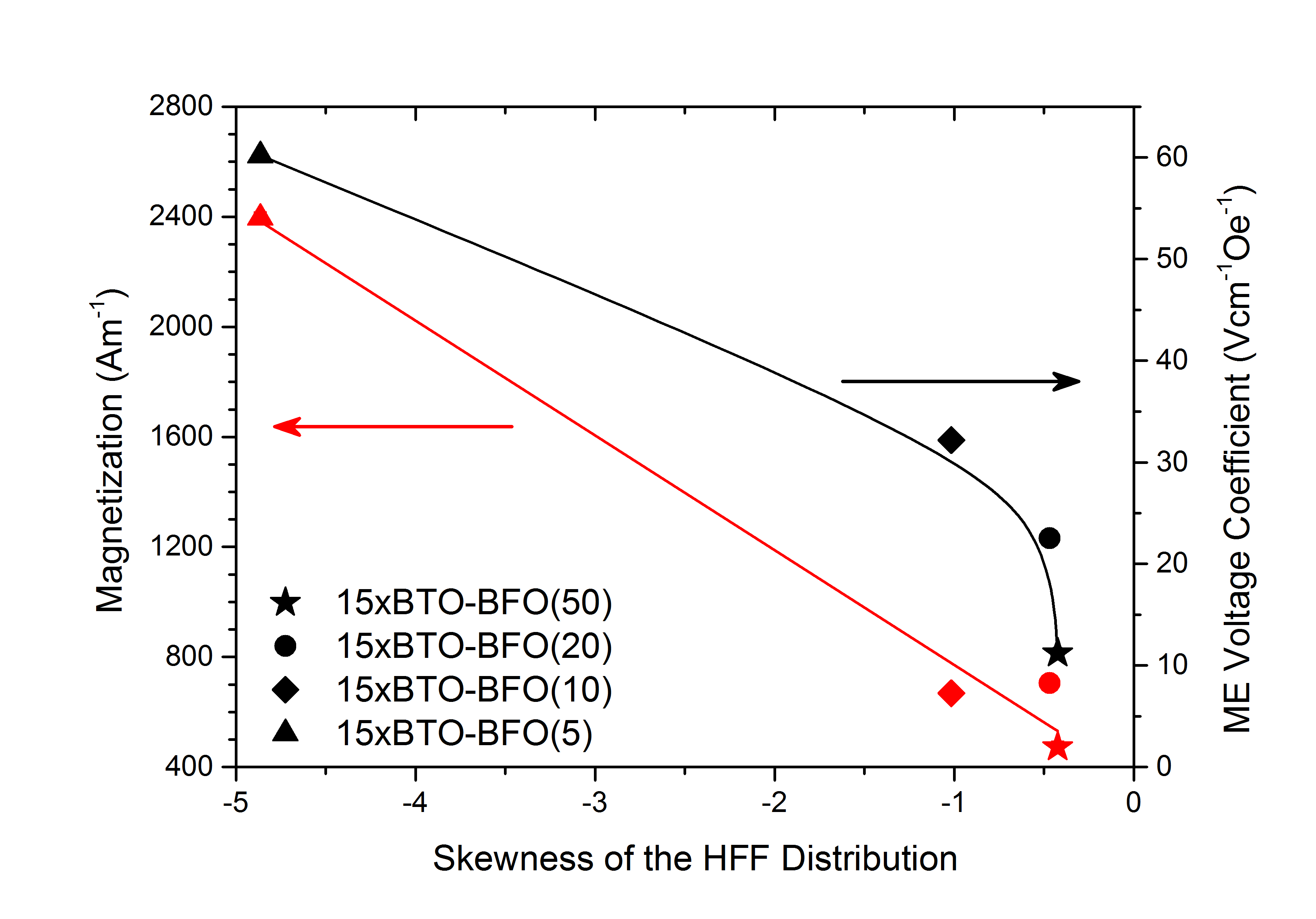}
	\caption{The relationship between magnetization, magnetoelectric voltage coefficient and the skewness of the hyperfine field (HFF) is shown (the straight lines are a guide to the eye). The skewness is a parameter describing the asymmetry of a distribution. The absolute value of the skewness increases as the asymmetry of a distribution increases. It can be seen that magnetization and ME voltage coefficient show the same trend with decreasing skewness.}
	\label{fig:correlation}
\end{figure}

To gain further insight into the magnetic behaviour of sample 15$\times$BTO-BFO(5) we probed the local magnetic environment of the $^{57}$Fe atoms in our multilayers using CEMS at room temperature. All four samples show the typical magnetic sextet (see Figure \ref{fig:Moessbauer}) with an asymmetry in the linewidth broadening of the first/sixth peak as is expected for iron in a position of trigonal symmetry which is the case for iron in BiFeO$_3$ \cite{Sando2013, Lebeugle2007}. 
The hyperfine field distributions of the samples (see Figure \ref{fig:Moessbauer}b)), however, do not feature the typical bimodal distribution known to occur in BiFeO$_3$ due to its spin spiral order \cite{Pokatilov2009}. 
This is an indication of the break-down of the spin spiral magnetic order, which leads to the emergence of different spin orders \cite{Pokatilov2009, Sando2013}. The HF field distribution of 15$\times$BTO-BFO(50) shows a symmetric peak around the average hyperfine field of 49.1~T. However, with decreasing thickness of the BiFeO$_3$ layer the distribution becomes more and more asymmetric with a tail towards lower fields. As a consequence, we can observe a decrease in average hyperfine field as a function of BiFeO$_3$ layer thickness. To quantify the degree of asymmetry of the hyperfine field distribution we use the weighted skewness $\sigma$ as defined in reference \cite{Rimoldini2014}. $\sigma$ is zero for a perfectly symmetric curve, while it assumes negative/positive values for negatively/positively skewed curves.
The macroscopic magnetization as well as the magnetoelectric coupling and the average hyperfine field strongly depend on the BiFeO$_3$ layer thickness.
In Figure \ref{fig:correlation} $\alpha_{ME}$ and the magnetization are shown as a function of the skewness of the HF field to point out the correlation between these three parameters. 
It can be clearly observed that $\alpha_{ME}$ as well as the magnetization increase with increasing asymmetry of the hyperfine field distribution.
This shows that the ME properties in BaTiO$_3$ - BiFeO$_3$ multilayers depend on the local structural and magnetic environment of the Fe ions in the crystal lattice.

The increasing asymmetry of the HF field distribution further implies that the magnetic environment of the $^{57}$Fe atoms becomes less symmetric, or less well ordered.
One contributing factor for a decrease in symmetry with decreasing layer thickness is the increasing surface - to - volume ratio. As the film thickness decreases the iron fraction that resides in an interface position of reduced symmetry increases, leading to a stronger contribution of the low symmetry iron.  
This effect is especially pronounced for sample 15$\times$BTO-BFO(5), where the asymmetric tail of the hyperfine field distribution extends down to zero Tesla.\\
In addition to the magnetic sextet a non magnetic doublet appears in the Mössbauer spectrum of 15$\times$BTO-BFO(5). This doublet originates from high spin paramagnetic $^{57}$Fe which might be due to the presence of some iron oxide precipitates. Possible evidence for such precipitates can be found in the STEM-EDX analysis of sample 15$\times$BTO-BFO(5) \cite{Lorenz2017}. It was found that the BiFeO$_3$ layers of sample 15$\times$BTO-BFO(5) contain up to 5~at.~\% of barium and titanium within the BiFeO$_3$ layers, which might indicate that up to 5~at.~\% of Bi and Fe atoms are substituted by Ba and Ti in the BiFeO$_3$ lattice. 
This intermixing of Ba and Ti into the BiFeO$_3$ layers of 15$\times$BTO-BFO(5) most likely leads to a lower leakage current and therefore increased polarization.
We found that the ME voltage coefficient increases as the percentage of low - symmetry iron increases. This is reflected in the increase of $\alpha_{ME}$ as a function of the skewness of the hyperfine field distribution.
\\ 
We summarize that an increased polarization and magnetization in combination with low mosaicity are important factors to explain the increase of $\alpha_{ME}$ in the investigated BaTiO$_3$-BiFeO$_3$ multilayer series. Furthermore, we discovered a strong correlation between the ME voltage coefficient, the magnetization and the symmetry of the hyperfine field distribution which indicates that for a better understanding of the magnetoelectric coupling the investigation of the local magnetic environment is essential.

BaTiO$_3$-BiFeO$_3$ multilayers with decreasing nominal BiFeO$_3$ layer thickness from 50~nm to 5~nm were grown and characterized. It was found that the magnetoelectric voltage coefficient strongly depends on the BiFeO$_3$ layer thickness. The largest value so far of 60.2 V cm$^{-1}$Oe$^{-1}$ was reached at 1~kHz, 10~Oe AC and 2~T DC magnetic field for the sample with a nominal BiFeO$_3$ thickness of 5~nm. The temperature dependencies of the magnetoelectric  voltage coefficient point to the conclusion that the main mechanism responsible for the magnetoelectric coupling at 300~K is dominated by charge transfer for the sample with a BiFeO$_3$ layer thickness of 50~nm. With decreasing BiFeO$_3$ thickness the main coupling mechanism is dominated more and more by strain. 
As the BiFeO$_3$ layer thickness decreases to 5~nm, the magnetization reaches its highest value, which exceeds the magnetization of the samples with a thicker BiFeO$_3$ layer by a factor of about two. 
EDX element distribution analysis done previously by our group \cite{Lorenz2017} showed a presence of about 5~at.~\% Ba and Ti in the BiFeO$_3$ layers of this specific sample, leading to an excess of iron in the BiFeO$_3$ layers. This result was confirmed by our CEMS measurements that show Fe oxide precipitates in the sample.
The Ba and Ti content in the BiFeO$_3$ layers most likely causes a decrease in leakage current and therefore an increase in polarization. 
The increased magnetoelectric coupling as well as the tunability of the magnetic and magnetoelectric properties of BaTiO$_3$-BiFeO$_3$ multilayers makes them an intriguing system. 

\section{Conflicts of Interest}
There are no conflicts of interest to declare.

\section{Acknowledgments}

We acknowledge the financial support from the Research Foundation Flanders (FWO), the Concerted Research Actions GOA/09/006 and GOA/14/007 of KU Leuven and the Hercules Foundation. Work at the University Leipzig was supported by the Deutsche Forschungsgemeinschaft within SFB 762 ``Functionality of oxide interfaces''.

\newpage
\section{Appendix - X-ray diffraction spectra and reciprocal space maps} \label{Appendix}

\begin{figure}[ht]
	\centering
		\includegraphics[width=0.75\textwidth]{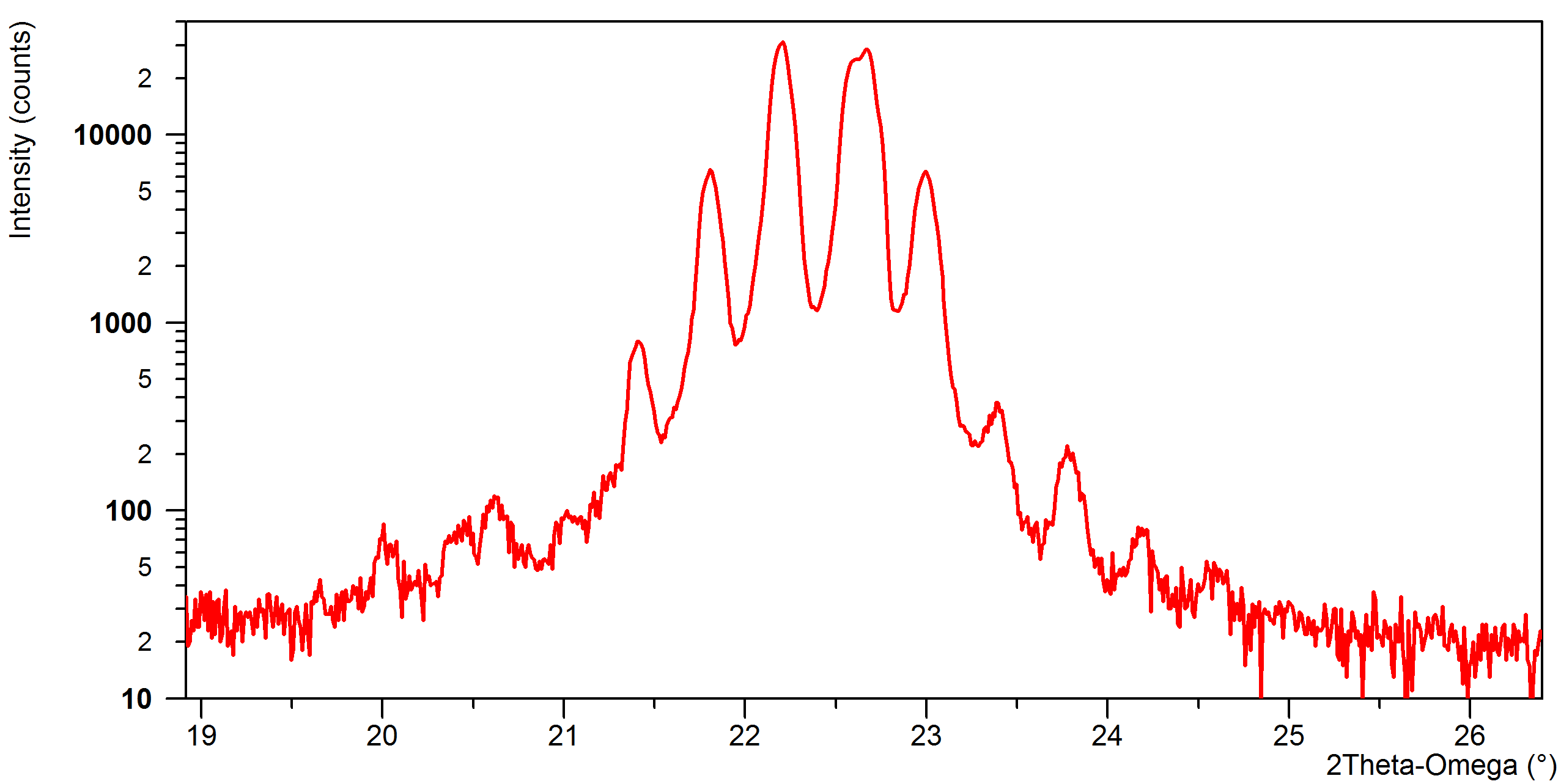}
	\caption{$2\theta$--$\omega$ scan of sample 20$\times$BTO-BFO, around the SrTiO$_3$(001) substrate peak which is the highest peak in center; mean fringe period: 0.39630$^{\circ}$, mean double layer thickness: 0.0232 $\pm$ 0.0003 $\mu$m}
	\label{fig:XRD_G5395}
\end{figure}

\vspace{2cm}

\begin{figure}[ht]
	\begin{adjustwidth}{}{}
	\centering
		\includegraphics[width=1.0\textwidth]{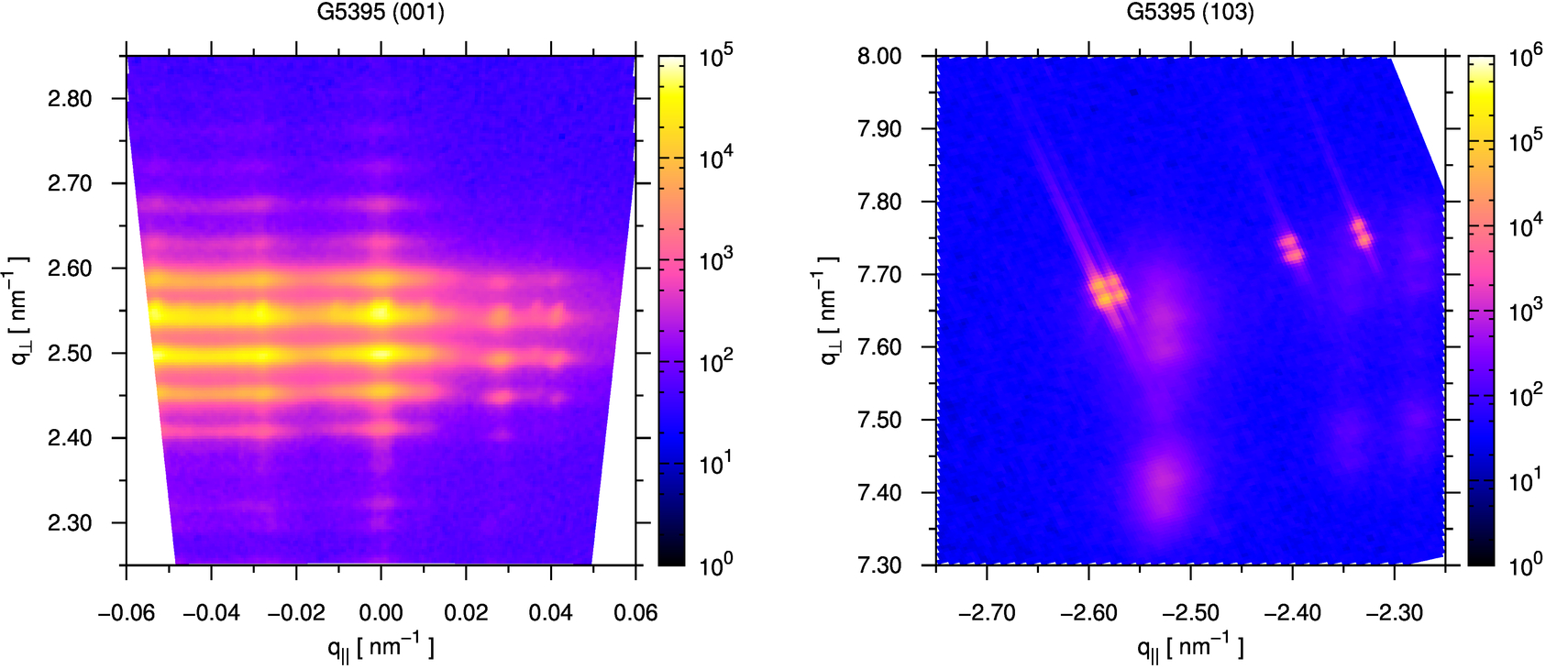}
	\caption{RSM of sample 20$\times$BTO-BFO on the left the symmetric (001) direction and on the right the asymmetric ($\overline{1}03$) direction}
	\label{fig:G5395_RSM}
	\end{adjustwidth}
\end{figure}

\clearpage

\begin{figure}[ht]
	\centering
		\includegraphics[width=0.75\textwidth]{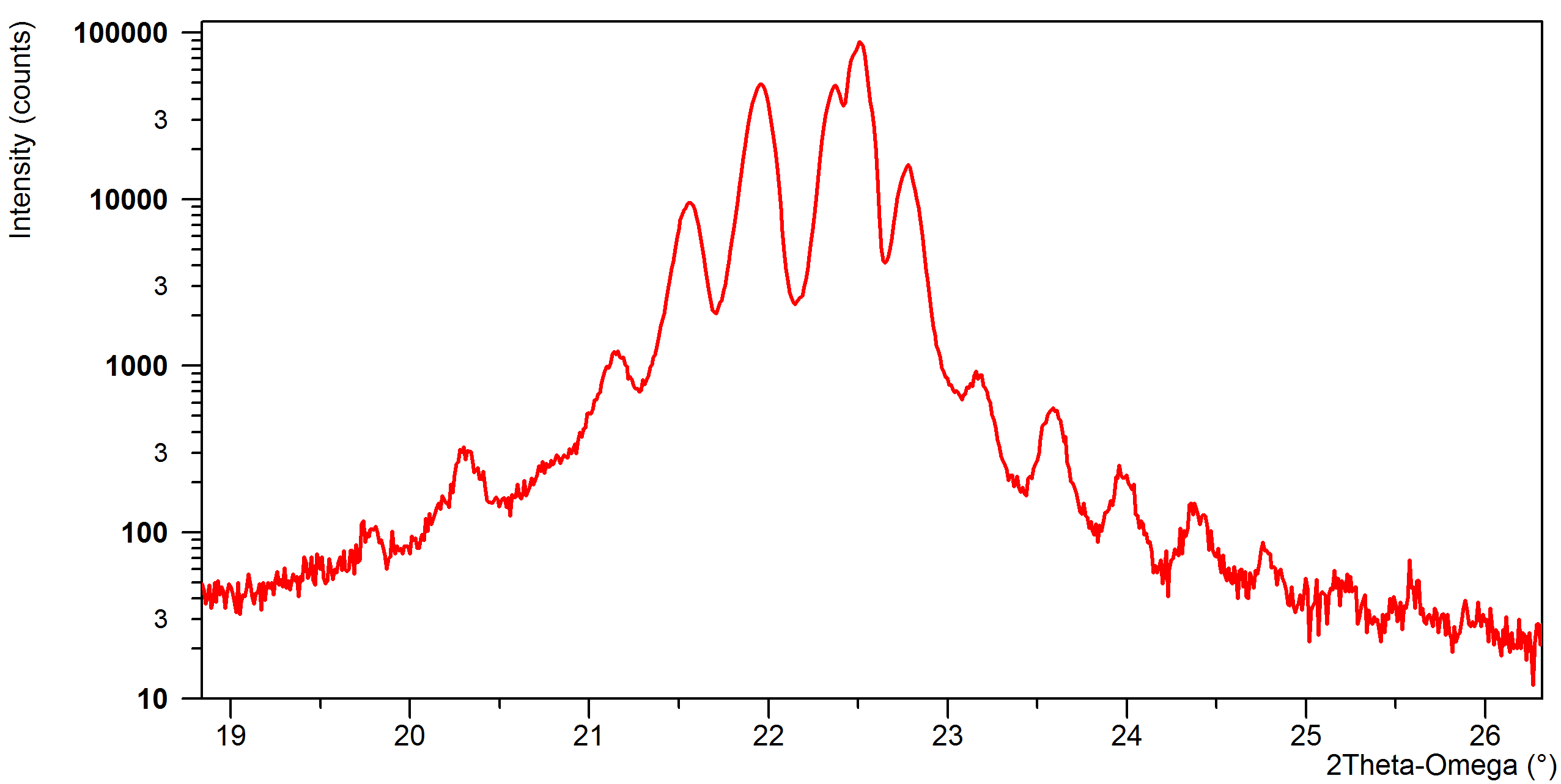}
	\caption{$2\theta$--$\omega$ scan of sample 10$\times$BTO-BFO, around the SrTiO$_3$(001) substrate peak which is the highest peak in center; mean fringe period: 0.40480$^{\circ}$, mean double layer thickness: 0.0221 $\pm$ 0.0006 $\mu$m}
	\label{fig:XRD_G5396}
\end{figure}

\vspace{2cm}

\begin{figure}[ht]
	\begin{adjustwidth}{}{}
	\centering
		\includegraphics[width=1.0\textwidth]{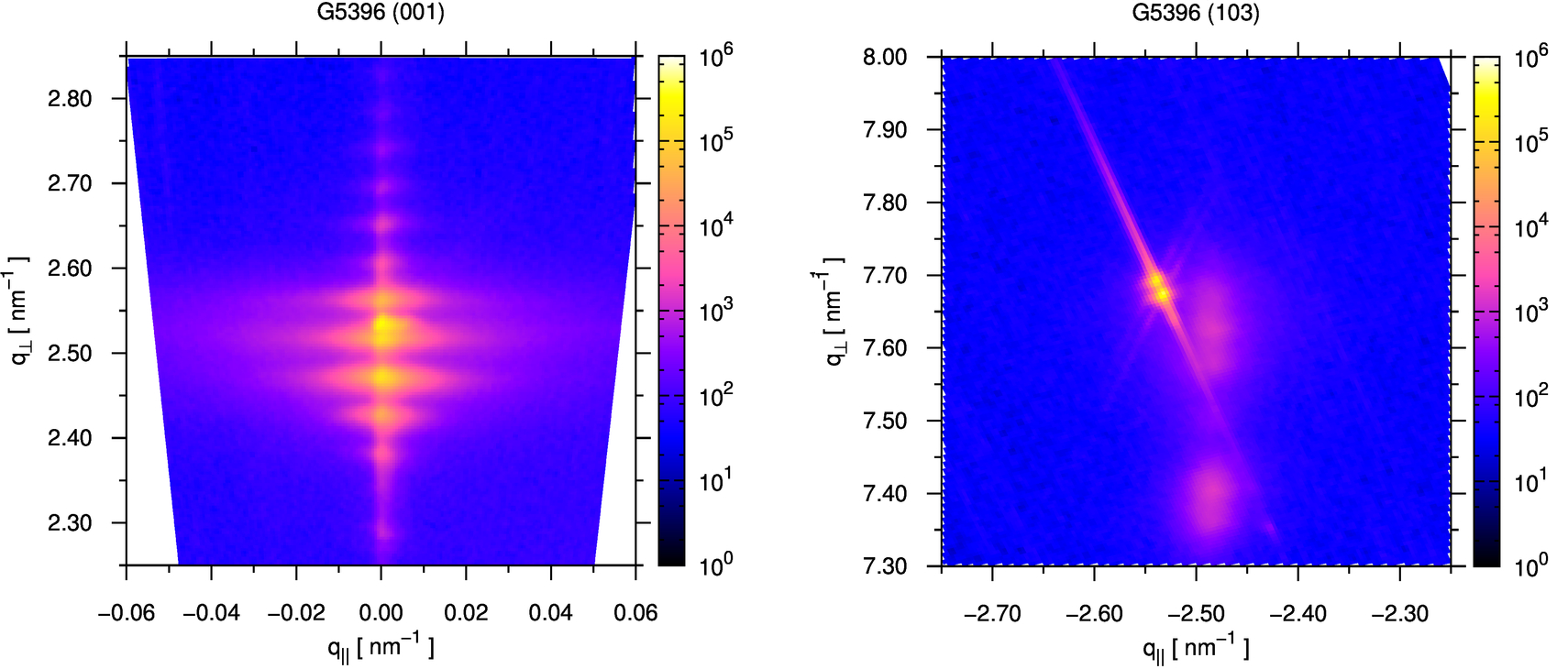}
	\caption{RSM of sample 10$\times$BTO-BFO on the left the symmetric (001) direction and on the right the asymmetric ($\overline{1}03$) direction}
	\label{fig:G5396_RSM}
	\end{adjustwidth}
\end{figure}

\clearpage

\begin{figure}[ht]
	\centering
		\includegraphics[width=0.75\textwidth]{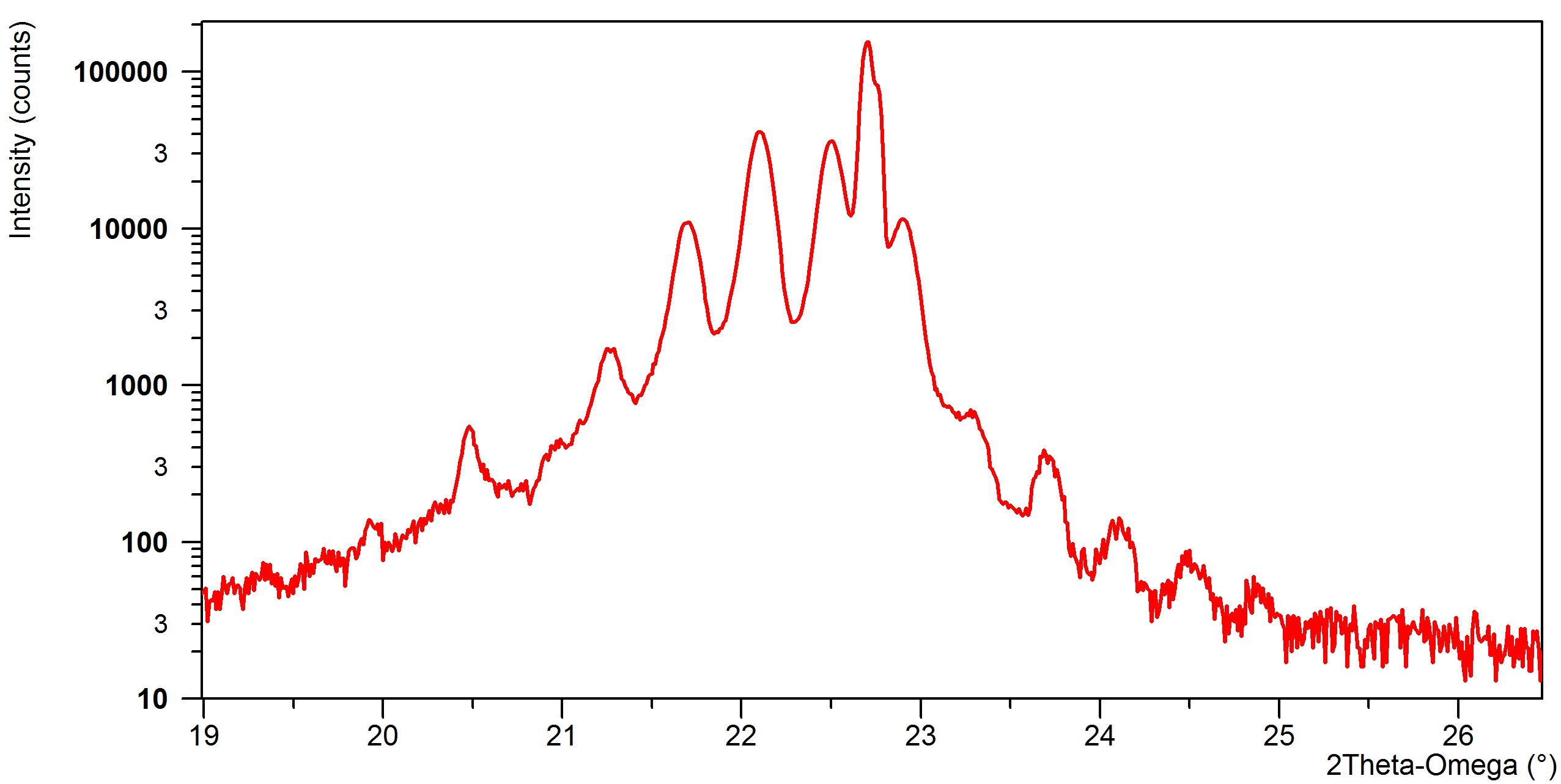}
	\caption{$2\theta$--$\omega$ scan of sample 5$\times$BTO-BFO, around the SrTiO$_3$(001) substrate peak which is the highest peak in center; mean fringe period: 0.40455$^{\circ}$, mean double layer thickness: 0.0228 $\pm$ 0.0006 $\mu$m}
	\label{fig:XRD_G5397}
\end{figure}

\vspace{2cm}

\begin{figure}[ht]
	\begin{adjustwidth}{}{}
	\centering
		\includegraphics[width=1.0\textwidth]{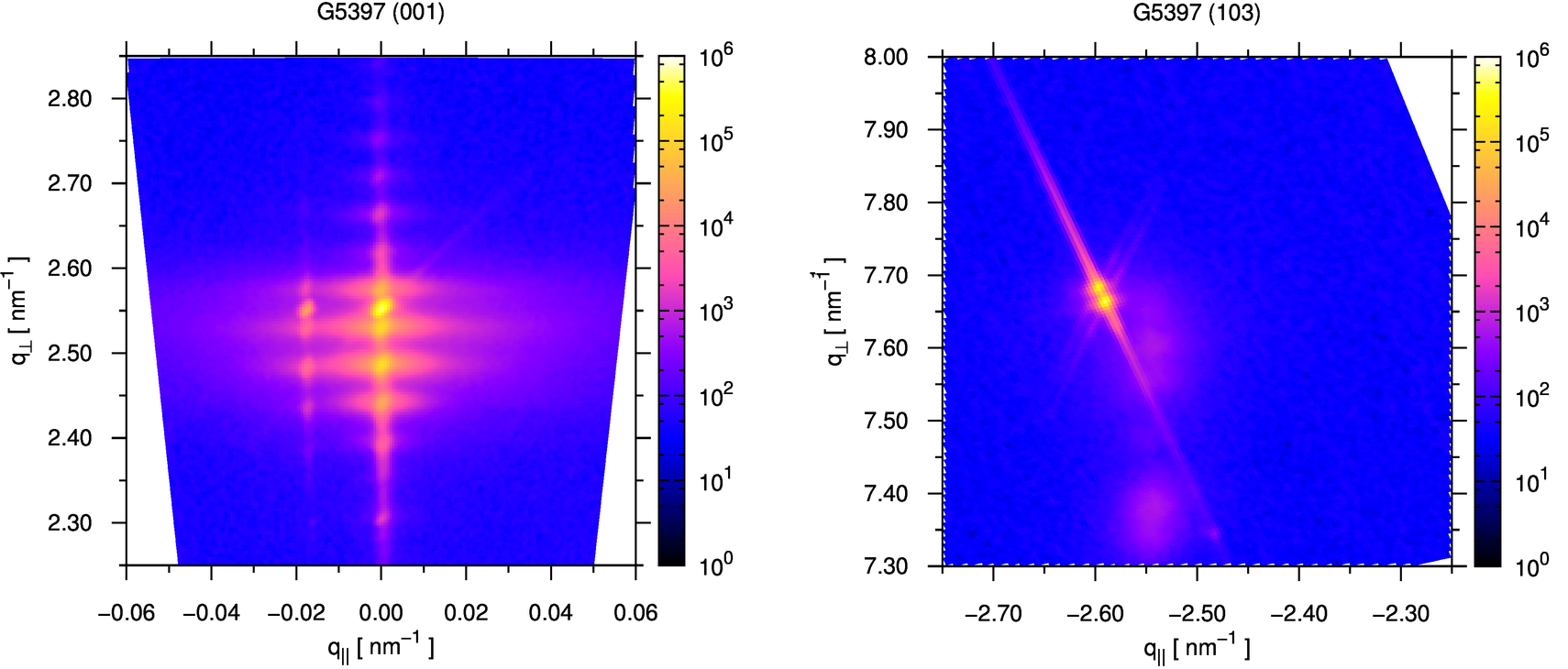}
	\caption{RSM of sample 5$\times$BTO-BFO on the left the symmetric (001) direction and on the right the asymmetric ($\overline{1}03$) direction}
	\label{fig:G5397_RSM}
	\end{adjustwidth}
\end{figure}

\clearpage

\begin{figure}[ht]
	\centering
		\includegraphics[width=0.75\textwidth]{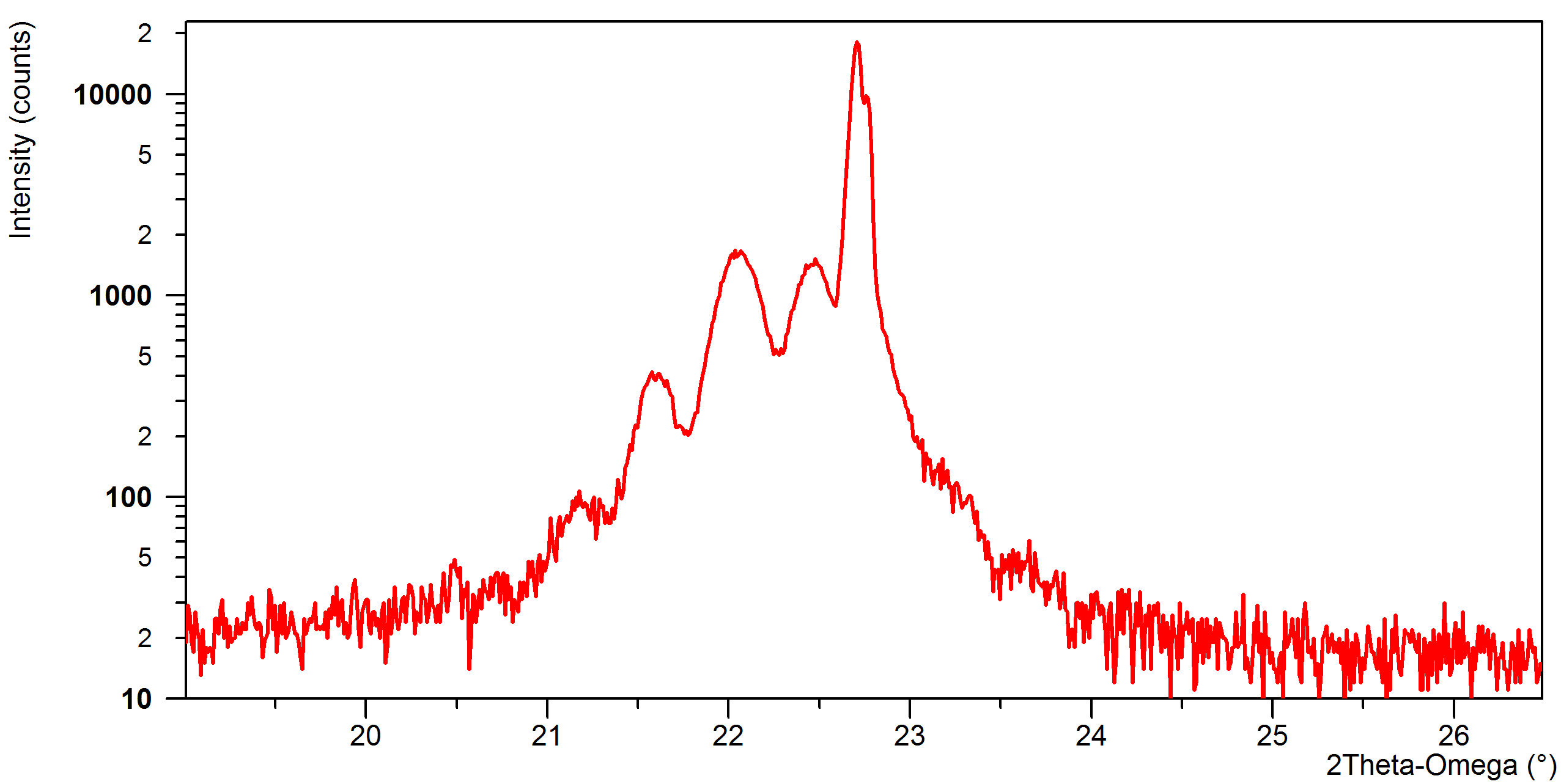}
	\caption{$2\theta$--$\omega$ scan of sample 2$\times$BTO-BFO, around the SrTiO$_3$(001) substrate peak which is the highest peak in center; mean fringe period: 0.42693$^{\circ}$, mean double layer thickness: 0.0211 $\pm$ 0.0008 $\mu$m}
	\label{fig:XRD_G5398}
\end{figure}

\vspace{2cm}

\begin{figure}[ht]
	\begin{adjustwidth}{}{}
	\centering
		\includegraphics[width=1.0\textwidth]{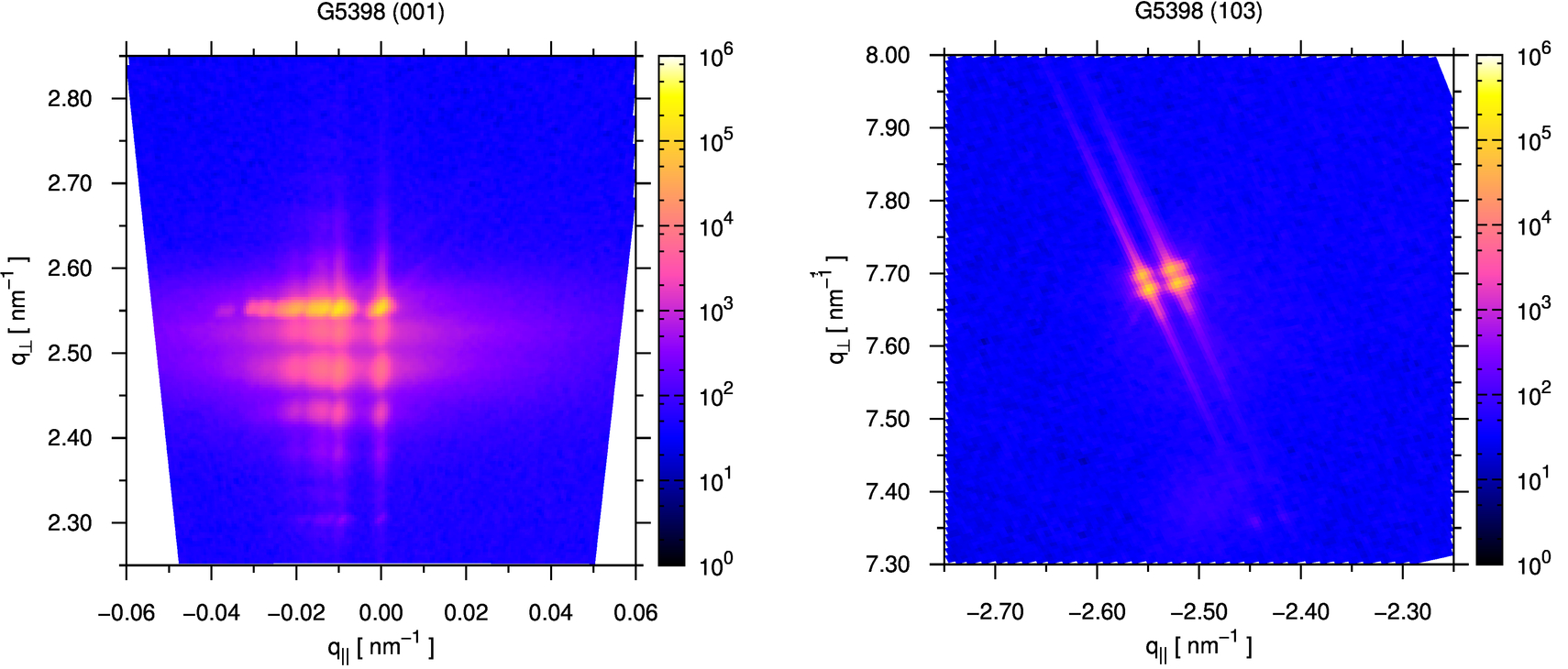}
	\caption{RSM of sample 2$\times$BTO-BFO on the left the symmetric (001) direction and on the right the asymmetric ($\overline{1}03$) direction}
	\label{fig:G5398_RSM}
	\end{adjustwidth}
\end{figure}

\newpage

\bibliographystyle{plain}
\bibliography{allpapers}

\end{document}